\documentclass[sigconf]{acmart}
\AtBeginDocument{%
  }

\setcopyright{acmlicensed}
\copyrightyear{2024}
\acmYear{2024}
\acmDOI{XXXXXXX.XXXXXXX}

\acmConference[CHI '25]{CHI Conference on Human Factors in Computing Systems}{April 26--May 1,
  2025}{Yokohama, Japan}

\begin{document}


\title{From Uncertainty to Innovation: Wearable Prototyping with ProtoBot}


\author{Ihsan Ozan Yıldırım}
\email{ihsanozan.yildirim@beko.com}
\orcid{0000-0002-2432-147X}
\authornotemark[1]
\affiliation{%
  \institution{Beko Corporate, R\&D Directorate, Sensor Technologies}
  \city{Istanbul}
  \country{Turkey}
}


\author{Cansu Çetin Er}
\email{cer20@ku.edu.tr}
\orcid{0000-0003-0957-3940}
\affiliation{%
  \institution{Koç University - Arçelik Research Center for Creative Industries (KUAR)}
  \city{Istanbul}
  \country{Turkey}
}


\author{Ege Keskin}
\email{ege.keskin@beko.com}
\orcid{0000-0002-5684-2229}
\affiliation{%
  \institution{Beko Corporate, R\&D Directorate, Sensor Technologies}
  \city{Istanbul}
  \country{Turkey}
}


\author{Murat Kuşcu}
\email{mkuscu@ku.edu.tr}
\orcid{0000-0002-8463-6027}
\affiliation{%
  \institution{Koç University, Electrical and Electronics Engineering Department}
  \city{Istanbul}
  \country{Turkey}
}


\author{Oğuzhan Özcan}
\email{oozcan@ku.edu.tr}
\orcid{0000-0002-4410-3955}
\affiliation{%
  \institution{Koç University - Arçelik Research Center for Creative Industries (KUAR)}
  \city{Istanbul}
  \country{Turkey}
}


\begin{abstract}

Despite AI advancements, individuals without software or hardware expertise still face barriers in designing wearable electronic devices due to the lack of code-free prototyping tools. To eliminate these barriers, we designed ProtoBot, leveraging large language models, and conducted a case study with four professionals from different disciplines through playful interaction. The study resulted in four unique wearable device concepts, with participants using Protobot to prototype selected components. From this experience, we learned that (1) uncertainty can be turned into a positive experience, (2) the ProtoBot should transform to reliably act as a guide, and (3) users need to adjust design parameters when interacting with the prototypes. Our work demonstrates, for the first time, the use of large language models in rapid prototyping of wearable electronics. We believe this approach will pioneer rapid prototyping without fear of uncertainties for people who want to develop both wearable prototypes and other products.

\end{abstract}


  





\maketitle

\section{Introduction}


Recently, AI, and especially large language models (LLMs), have rapidly begun to permeate daily life and simplify tasks across various fields \cite{wu2023brief}. Furthermore, AI assistants are increasingly being utilized in design and production processes in areas such as software development, user-interface design, computer-aided design, and education \cite{petridis_promptinfuser_2024, sabuncuoglu_prototyping_2022}. Similarly, designing wearable technologies involves processes like software and hardware design, which are increasingly benefiting from AI support. However, creating prototypes of wearable electronic devices without expert assistance in these areas remains a challenging endeavor. To the best of our knowledge, there is no tool that enables users with limited software and hardware knowledge to prototype without the need to write code. This is significant because designing wearables is a complex task that encompasses multiple aspects, such as interactivity, functionality, and social context \cite{genc_toolkits_2022}.



In wearable technology design research, play and game contexts are frequently employed, particularly within the scope of embodied play \cite{buruk_design_2019, mehta_arm--dine_2018, jorgensen_body_2024}. According to the framework developed in this field, exploring wearables through the lens of playful interaction provides opportunities to enhance self-expression, create diverse social dynamics, and define various interaction modalities \cite{buruk_design_2019}. Additionally, situated play design is utilized to capture playful moments in daily non-entertainment-based activities and transform them into design opportunities \cite{altarriba_bertran_chasing_2019}. Following this approach, we conducted our study using a case study methodology to enable users to easily explore their daily life experiences and receive rapid feedback.


To address this gap from the perspective of smart wearables and playful interactions, we designed ProtoBot, an AI chatbot that allows users to prototype wearable devices through prompt-based interaction, without the need to know how to code directly ProtoBot is a web-based chatbot developed using Python and Flask, intended for use with Arduino-compatible and microcontroller-based wearable electronic prototyping boards—such as Lilypad\cite{buechley2008lilypad}, Adafruit Flora\cite{adafruit_flora}, and Deneyap G\cite{butuner_deneyap_2021, deneyap_kart_g_type_c}—as well as prototyping kits. It leverages the OpenAI ChatGPT API\cite{openai_api_reference} and the Arduino CLI\cite{arduino_cli} to facilitate the design of interactions and wearable devices using sensors and actuators.

In this study, we utilized the design thinking-based iterative sensor and actuator selection flow \cite{ozan_yildirim_concept_2024} for the selection of sensors and actuators while developing prototypes with Deneyap G boards and sensors and actuators from Deneyap Prototyping Kits (Deneyap Card Robotics Programming Set, Deneyap Card Smart Home Set) \cite{butuner_deneyap_2021}. We conducted a case study with four professionals from different disciplines of psychology, business, project management, and interaction design research. As a result of this study, four distinct wearable device concepts emerged. To realize these concepts, we asked the participants to break down their ideas into smaller components and create the desired prototypes and interactions using ProtoBot. Through the process of creating these prototypes, we learned three design lessons about the user experience provided by this tool and how it can be improved: (1) we observed that the uncertainty inherent in such a development process was transformed into a positive experience, (2) the role of the chatbot should act as a guide for users with limited knowledge, and (3) users need to adjust design parameters when interacting with the resulting prototypes.

Our contribution is two-fold. First, by exploring four design concepts, we have successfully demonstrated that ProtoBot functions seamlessly and observed its positive impact on reducing the anxiety associated with inexperience in hardware and software design. Secondly, we have derived three lessons regarding participants' experiences and tool design for future studies. We believe that our work lays the foundation for helping individuals with limited hardware and software knowledge to explore wearable prototyping possibilities.

\section{Background}

Designing wearable electronic devices traditionally demands specialized knowledge in software and hardware, posing significant challenges for individuals without this expertise. Recent developments aim to lower these barriers, enabling more people to engage in wearable technology design and prototyping. Two key areas contributing to this democratization are wearable prototyping kits, which offer modular and user-friendly platforms for device creation, and playful wearables, which utilize engaging interactions to enhance user experience and stimulate creativity. In this section, we explore relevant work in these areas to provide context for ProtoBot's development and its role in simplifying the prototyping process for non-expert users.

\subsection{Wearable Prototyping Kits}

The development of modular and flexible prototyping kits has been a significant focus in the field of wearable technology design. These kits aim to provide designers with the ability to mix smart and non-smart components while maintaining functionality in a modular fashion. One common approach to achieve modularity is through the use of interconnected modules forming a skeletal structure. Connections between these modules can be electrical—for communication or power—or structural, to maintain the integrity of the design.

Several toolkits have been developed following this approach, offering varying degrees of modularity. Notable examples include the Lilypad \cite{buechley2008lilypad}, Adafruit Flora \cite{adafruit_flora}, SoftMod \cite{lambrichts2020softmod}, and Snowflakes \cite{buruk2021snowflakes}. These kits allow designers to experiment with different configurations and functionalities, facilitating creative exploration in wearable design. Additionally, other kits like LittleBits \cite{littlebits} and Wearable Bits \cite{jones2020wearable} have contributed to advancing modular systems in wearable prototyping.

An ongoing effort in this emerging field is the seamless integration of functionality with form. Depending on the design objectives and the desired user experience, designers may choose to either highlight or conceal smart components. Toolkits such as BLInG \cite{jo2021bling}, SkinKit \cite{ku2021skinkit}, and Wearable Bits \cite{jones2020wearable} provide options for integrating smart components either visibly or discreetly, allowing for greater flexibility in design aesthetics. Despite the availability of these toolkits, there remains a gap in tools that enable designers to creatively explore options for hiding, revealing, and resizing smart components while mixing various materials and colors. Previous works like Snowflakes \cite{buruk2021snowflakes} have made significant strides in this area but still have limitations in terms of size and form.


By leveraging such prototyping kits, designers can more easily experiment with wearable technologies, integrating functionality and form in innovative ways. However, there is still a need for tools that simplify the prototyping process for users with limited software and hardware knowledge, which is where ProtoBot aims to make a significant contribution.

\subsection{Playful Wearables}

Playful wearables have applications in various fields, such as enhancing immersion in pervasive play \cite{waern_three-sixty_2009}, acting as mediators in social interactions \cite{buruk_design_2019, mehta_arm--dine_2018}, and allowing users to visually transform into fictional characters during role-playing games \cite{jing_magia_2017}.

Beyond these uses, playful interactions in wearables hold broader potential to make an impact spanning from the urban scale to the homes and individuals. As an instance of the urban context, \textit{Self-Sustainable Chair} \cite{papadopoulos_wearable_2007} acts as a wearable dress that transforms into a chair through user movement to help users rest and requires walking to re-inflate while strolling in the city. Another example is \textit{High Water Pants}, which is a mechatronic pair of pants used by everyday cyclists that gives tangible feedback during cycling to create awareness of climate change.

Moreover, wearables also enrich spatial and individual experiences. One speculative example is a coatrack that changes its function to facilitate communication and social interaction \cite{oogjes_designing_2018}. Each time a home member comes home and hangs up their coat on this rack, the digital connection at the home gradually decreases, allowing the home members to share more intimate and sentimental moments when they are physically together. Many other examples involve wearables that enhance individual experiences, either shared or solitary. Among these, \textit{Arm-A-Dine} showcases an augmented dining experience where two users feed each other using robotic arms.

In this case study, we used ProtoBot to further explore how individuals with limited hardware and software knowledge can ideate on generating individual playful experiences with wearable electronics without the limitation of the knowledge barrier.

\section{Method}


This study employed a design thinking-based iterative approach for selecting sensors and actuators, as outlined in \cite{ozan_yildirim_concept_2024}. Utilizing Deneyap G \cite{deneyap_kart_g_type_c} boards alongside sensors and actuators from Deneyap Prototyping Kits (Deneyap Card Robotics Programming Set \cite{deneyapkart2024robotikprogramlama}, Deneyap Card Smart Home Set \cite{deneyapkart2024akilliev}), participants developed prototypes through a structured process. Each prototyping session was designed to last approximately one and a half hours, providing ample time for participants to engage deeply with the tools and materials while developing their wearable device concepts.

\subsection{Procedure}

Each session consisted of two moderators and a single participant. One of the moderators actively engaged with the participant during the study, while the other conducted the post-study interview. All sessions, including the post-questionnaire procedures, were carried out in the participants' native language, Turkish.

At the beginning of each session, participants received a brief introduction to the purpose and scope of the study. Afterwards, the sensor and actuator modules were revealed, and their functions were summarized by the moderator. The participant was then prompted to develop a playful, wearable concept using the sensor and actuator modules. Following \cite{ozan_yildirim_concept_2024}, participants were also encouraged to consider the \textit{affordances and physical phenomena} prior to selecting sensor and actuator modules. 

After proposing their concept and identifying the necessary modules, participants were asked to connect them using the provided standard wires. For simplicity, all connectors and wiring were standard and identical, utilizing I2C connectors and cables known as Sparkfun Qwiic (Quick I2C) \cite{Bell2021}, Adafruit Stemma QT, or the Grove System. This standardization ensured that participants were not required to identify specific connector types. Additionally, the modules were daisy-chained, further simplifying the construction of the prototype.

Once the electrical connections were completed, participants were instructed to interact with ProtoBot to upload the code. They were asked to briefly describe their prototype's components and desired functionalities. As ProtoBot generated the code, participants experimented with attaching the prototype to their bodies. When the code was ready, it was uploaded to the prototype using ProtoBot. Participants were then asked to wear the prototype, using provided tapes and rubber bands for attachment. At this stage, moderators might partially intervene to assist with the process. If participants could easily alter the readings of the sensor(s) (e.g., changing the temperature sensor value by holding the device), they were encouraged to use the prototype in a simulated real-world scenario. The session concluded once participants had a general understanding of how the prototype felt while using it.

The post-study interview was structured qualitatively around three main questions. First, participants were asked about their expectations prior to the study and how the actual experience aligned with those expectations. We aimed to learn what they gained from the experience, what challenges they encountered, and what could be further improved. Next, we explored how the prototyping process could have been done differently. This included discussions on how they might rewrite the prompts after the experience, what other interactions they could imagine, and which additional sensors they might consider using for these new interactions. Finally, to conclude the interview, participants were asked how their thoughts on prototyping wearables with electronic components had changed after the experience and how they could apply this knowledge in their own fields of study or personal interests.

\subsection{Participants}

We recruited four professionals from diverse disciplines to participate in our study. The selection aimed to encompass a range of backgrounds and expertise levels related to software and hardware design, providing insights into ProtoBot's usability for non-expert users.

Participant 1 (P1): A communications expert specializing in the energy market industry. P1 has limited experience with hardware and software prototyping but possesses a keen interest in technology and its applications in business contexts. Participant 2 (P2): An interaction design researcher specializing in designing interactive systems. While P3 has theoretical knowledge of user experience principles, they have limited experience with the practical aspects of hardware implementation. Participant 3 (P3): A project manager with exposure to technology projects but without hands-on experience in prototyping wearable devices. P2 is interested in how emerging technologies can streamline project workflows and enhance team collaboration. Participant 4 (P4): A psychologist interested in human-computer interaction and the psychological aspects of technology use. P4 has no technical expertise in electronics or programming but is curious about the potential of technology to impact well-being and behavior.

All participants were volunteers who expressed interest in exploring wearable technology through ProtoBot. They had no prior experience with ProtoBot or similar AI-driven prototyping tools. By engaging professionals from different disciplines with varying levels of technical expertise, we aimed to assess the effectiveness of ProtoBot in enabling users without coding knowledge to prototype wearable devices. This diversity allowed us to gather a broad range of feedback on the tool's usability, accessibility, and potential areas for improvement.

\section{ProtoBot}

ProtoBot is a chatbot concept developed using the Python programming language, designed to enable individuals with limited software and hardware expertise to easily program Arduino-compatible devices. This chatbot leverages the OpenAI ChatGPT-3.5 \cite{wu2023brief} model through the OpenAI ChatGPT API \cite{openai_api_reference}, utilizing the OpenAI Python library \cite{openai2024openai-python} for backend operations. The frontend of ProtoBot is developed with Python Flask, providing a user-friendly web interface. ProtoBot generates code in the Arduino programming language, a variant of C and C++, and facilitates the uploading of this code to Arduino-compatible devices via the Arduino Command Line Interface (CLI) \cite{arduino_cli}. Users interact with ProtoBot through prompt-based text conversations, allowing them to create and modify wearable device prototypes without requiring direct knowledge of coding. The current source code for ProtoBot is accessible through the ProtoBot GitHub Repository \cite{protobotgithub}.

To ensure that ProtoBot generates only relevant and executable code, it utilizes predefined system prompts when interacting with the ChatGPT-3.5 model. An example of such a prompt is: 

\texttt{[{"role": "system", "content": "You are an expert Arduino programmer. Only return valid and complete Arduino code, without any explanations or comments."}]}

End users are not informed about the specific LLM model ProtoBot employs or the details of its fine-tuning process, ensuring a seamless and focused user experience.

\begin{figure}
    \centering
    \includegraphics[width=0.5\linewidth]{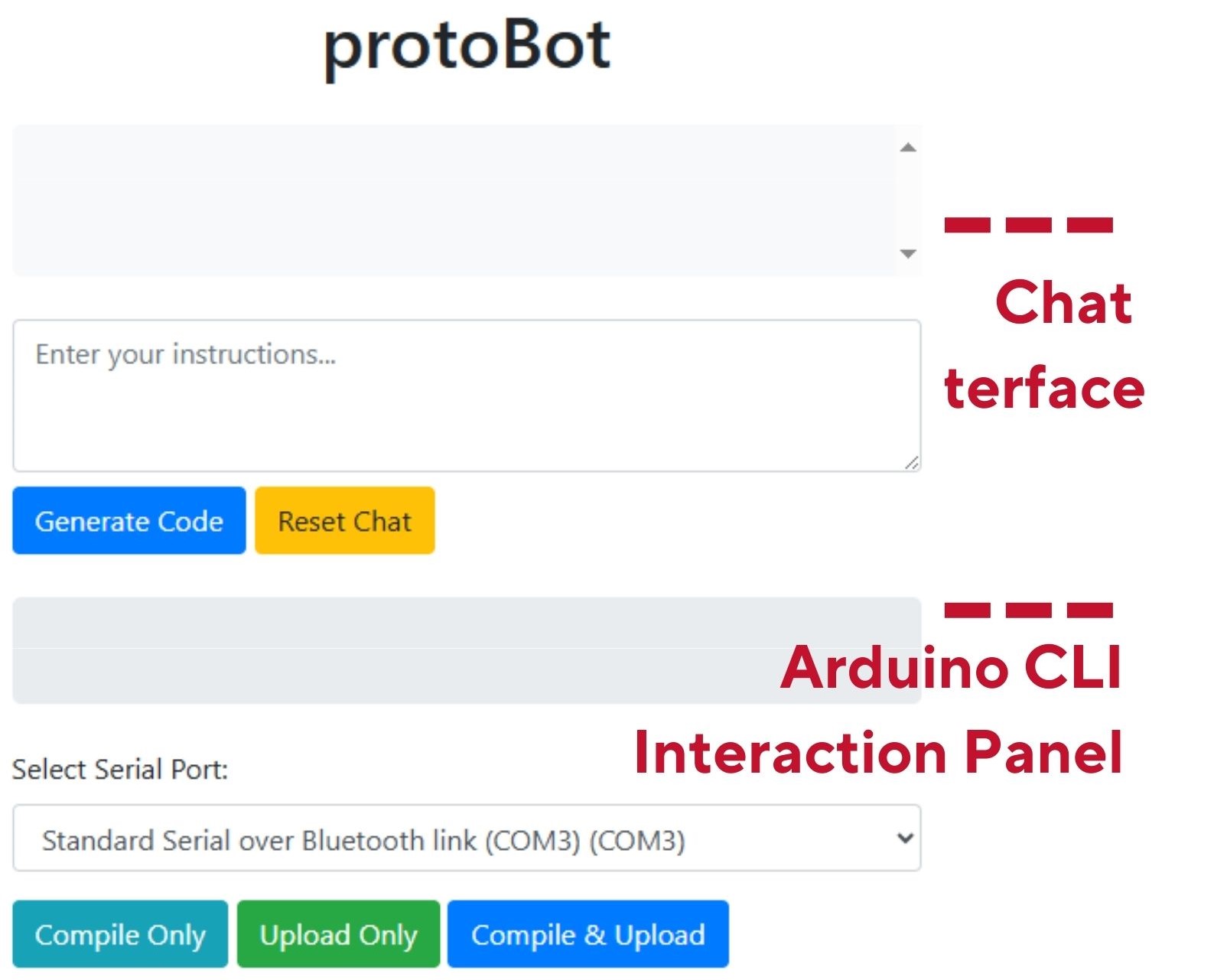}
    \caption{ProtoBot User Interface}
    \label{fig:protobotui}
    \Description{ProtoBot User Interface}
\end{figure}

The ProtoBot interface comprises two main components: a chat interface and an Arduino CLI Interaction Panel. (see Figure~\ref{fig:protobotui}): 

\textbf{Chat Interface}: This section facilitates continuous text-based conversations between the user and the chatbot. It includes a chat history area where previous interactions are displayed and an instruction input field where users can enter their prompts or queries.

\textbf{Arduino CLI Interaction Panel}: Located below the chat interface, this panel becomes active once ProtoBot successfully generates code. It displays the generated Arduino code and provides an interface for uploading the code to connected devices via the Arduino CLI. Users are prompted to select the device connected to their computer, and they are offered options to compile the code, upload it to the device, or perform both actions simultaneously. Additionally, the output of the upload process is displayed at the bottom of the interface, allowing users to review the results and continue the conversation with ProtoBot by providing new instructions based on the output.

In this study, the version of ProtoBot utilized was specifically developed to support Deneyap G boards. All libraries for the modules included in the Deneyap prototyping kits were loaded onto the computer running ProtoBot via the Arduino CLI, ensuring seamless integration and functionality during the prototyping process.

\subsection{Electronic Components}

ProtoBot is designed to be compatible with all Arduino-based development boards, offering flexibility for various wearable electronics projects. However, to streamline the development process and minimize hardware setup efforts within the scope of this study, we employed Deneyap G boards alongside sensors and actuators from Deneyap's Prototyping Kits, including the Deneyap Card Robotics Programming Set and Deneyap Card Smart Home Set. These kits support the use of standard I2C connection cables \cite{Bell2021}, facilitating easy integration of multiple sensors and actuators.

\begin{figure}[H]
    \centering
    \includegraphics[width=0.8\linewidth]{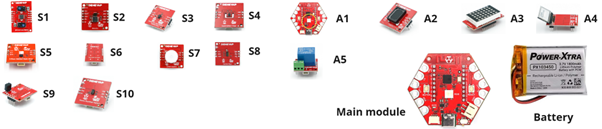}
    \caption{Sensors used in the case study: S1 - Line Following Sensor, S2 - Microphone, S3 - APDS-9960 Digital RGB, Ambient Light, Proximity, and Gesture Sensor, S4 - SHTC3 Temperature and Relative Humidity Sensor, S5 - LSM6DSM IMU, S6 - MSP430G2352 Touchpad, S7 - PIR Sensor, S8 - VL53L0CXV0DH/1 Time of Flight Sensor, S9 - IR Transmitter and Receiver, S10 - LTR-553ALS-01 Ambient Light Sensor. Actuators and main components used in the case study: A1 - Addressable RGB LED (on main module), A2 - Speaker, A3 - 5x7 LED Matrix, A4 - OLED Display, A5 - Relay, Main Module - Deneyap G, Battery - 3.7V 1800mAh Lithium Polymer.}
    \label{fig:sensors}
    \Description{Sensors and actuators used in the case study}
\end{figure}

Participants were introduced to the sensors and actuators available within these kits through a presentation page (see Figure~\ref{fig:sensors}). This allowed them to understand the capabilities and limitations of each component used in their prototypes. For detailed information about the sensors and actuators, participants were directed to the prototyping kits' official web pages \cite{deneyapkart2024robotikprogramlama,deneyapkart2024akilliev} and the Deneyap GitHub account \cite{deneyapgithub}.


By utilizing Deneyap G boards and their associated modules, ProtoBot provided a robust platform for developing and testing wearable electronic prototypes. This setup ensured that participants could focus on the design and interaction aspects of their projects without being hindered by complex hardware configurations.




\section{Concepts Generated During the Case Study}




\begin{figure}
    \centering
    \includegraphics[width=0.7\linewidth]{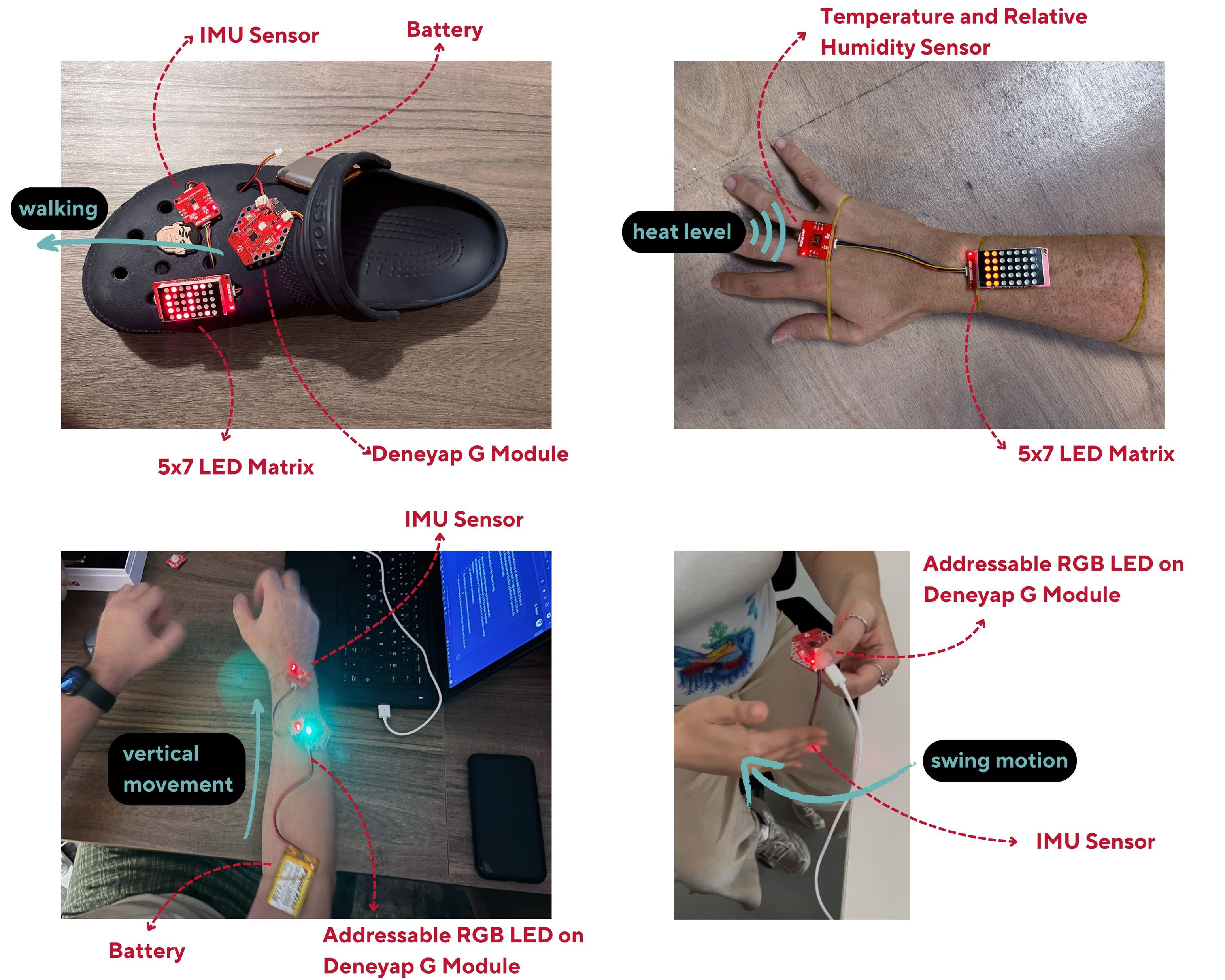}
    \caption{Wearable device concepts generated by participants: \textit{GuidingSteps} (upper left), \textit{PedalPulse} (upper right), \textit{BakeHero Glove} (lower left), and \textit{FitFit} (lower right).}
    \label{fig:conceptPhotos}
    \Description{Wearable device concepts generated by participants, including GuidingSteps, PedalPulse, BakeHero Glove, and FitFit.}\end{figure}

\textbf{Concept 1, GuidingSteps: Footwear that guides each step you take} In this case, P1 ideated a navigation system that could be used while traveling abroad. The navigation system would be embedded in shoes or slippers, allowing users to find their way without constantly checking their phones. This interaction reminded the participant of a scavenger hunt through an unfamiliar city.


After physically prototyping the concept that can be seen in \ref{fig:conceptPhotos}, the participant provided the following prompt and attempted to run the system. After a few necessary iterations, they observed that the system functioned similarly to what they had envisioned. Initial prompt: "\textit{We have a Deneyap G board, to which we have connected a 5x7 matrix LED, as well as an I2C IMU (LSM6DSM). Could you create a simple navigation experience code for us? We will also add a battery to this setup and attach it to our shoe. For example, you could create a route that provides guidance like 'go straight for 10 steps' and 'turn right for 5 steps.'"}

\textbf{Concept 2, BakeHero Glove: Encourages quick action to 'beat the heat!'} For this scenario, P2 focused on the kitchen context. The participant highlighted that the kitchen is a complex space where cooking demands attention due to various stimuli such as sound, heat, and humidity. Among these activities, P2 decided to gamify the process of removing muffins from the oven without getting burned. They suggested that if a game-like bar resembling flames could be seen, it would encourage quicker and more fun action.



After creating a physical prototype that can be seen in \ref{fig:conceptPhotos}, P2 provided the following prompt and attempted to run the system. Initial prompt: \textit{"Hello, I have connected a Deneyap G board to my computer with a usb cable. It also has a temperature sensor named "Deneyap Sicaklik nem ölçer". I have also attached a 5x7 led matrix named "5x7 Led Matris", via the I2C connectors. I want to wear this module on my hand, and I want it to measure the ambient temperature near my hand. And based on the temperature value, I want the rows of the LED matrix to light up as if it is a bar indicator. I don't want my hand to get too hot, I want to use this system as a visual warning for my hand, so that I can prevent myself putting my hand into a very hot oven or near an open fire or a stove. Can you provide the code for this operation?"}

\textbf{Concept 3, PedalPulse: Provides visual feedback on cycling intensity} In this case, P3 imagined using wearables in the sports context. The participant mainly focused on receiving feedback to let them know when to change gears or remind them to get hydrated after a certain distance. P3 saw this as an opportunity to challenge their cycling performance. After considering different aspects of cycling, the participant decided to focus on calculating acceleration through changes in inclination with a gear worn around the arm.



After building a physical prototype that can be seen in \ref{fig:conceptPhotos}, P3 provided a prompt and tested the system. With a few necessary iterations, they observed that the system functioned similarly to what they had envisioned. Initial prompt: \textit{"I have a Deneyap G board which has an addressable RGB LED on it. Could you write me the code that changes the colors of the RGB LED based on the inclination information? We have also connected an I2C IMU sensor (LSM6DSM) to measure the inclination on the device."}

\textbf{Concept 4, FitFit: Bringing playful awareness to every paw movement} For this concept, P4 explored more-than-human interaction when thinking from the perspective of playful interactions. The participant explained that they would have a closer connection to their cat if they were more aware of the cat's interactions. She imagined two connected devices, worn by herself and the other by her cat, that would provide feedback after the cat used the litter box. The participant’s device would replicate the "FitFit" noise the cat made during the interaction through an earring, which almost acts as a headphone.




After physically prototyping the concept that can be seen in \ref{fig:conceptPhotos}, the participant provided the following prompt and attempted to run the system. After a few necessary iterations, they observed that the system functioned similarly to what they had envisioned. Initial prompt: \textit{"I have a Deneyap G board which has an addressable RGB LED on it, and I attached an IMU sensor (LSM6DSM) to it via the provided I2C cable. I attached this setup to the belly of my cat to measure of his paws when he is burying his poop. Could you please generate the code to count how many times he used his paws when this process happens and give feedback using RGB led. For example 50 times of movement will be given as a red light on the sensor as a feedback."}

\section{Lessons Learned \& Opportunities for Future}

\subsection{Turning Uncertainty Into a Positive Experience}

One of the key observations from our study was how participants transformed initial feelings of uncertainty into a positive and motivating experience. P4 stated that at the beginning of the study, she felt anxious upon seeing many unfamiliar electronic components. After the study process, she remarked, \textit{"The initial uncertainty gradually decreased and eventually turned into a positive uncertainty for me. After playing with these components like a game, there is still some uncertainty, but now it makes me feel excited about what else I can do with these."} Similarly, P2 mentioned that becoming familiar with the components led him to think from a minimalist perspective.

As a result of this study, we observed that even with ProtoBot—which is itself still a prototype—the common feeling of "I can't do this. It requires a lot of knowledge on many things" regarding making electronic prototypes can be quickly overcome. Along with this potential, we found that ProtoBot has the ability to facilitate the design process for designers and to encourage non-design professionals to create designs, even in our work using electronic components that can be further developed in terms of wearability.

Moving forward, as ProtoBot and similar tools are developed, we foresee that it will be possible to create instant working prototypes through a written or verbal conversation where the user only needs to share whether it worked as desired or not, without needing to know that code is being developed in the background. As AI algorithms continue to advance, we predict that tools like ProtoBot will easily enable the prototyping of systems involving multiple interactions. In this study, we developed a tool intended for wearable prototyping; however, we realized that when developing tools like ProtoBot, they can actually serve as a conceptual framework that addresses the entire end-to-end design process.

\subsection{ProtoBot Acting as a Reliable Guide}

An important aspect we observed was ProtoBot's potential to act as a reliable guide for users with limited expertise. P2 realized that, based on his prior experience with AI chatbots, inconsistencies in the prompts he provided to ProtoBot caused the AI to misunderstand his instructions. When he had to iterate several times, he observed that the code output generated by ProtoBot, upon being uploaded to the electronic components, produced results different from what was requested. As a suggestion, he mentioned that ProtoBot could become a more reliable guide if more consistent data were provided. He remarked, "For example, ProtoBot could ask me to upload photos of the two components I have and write three sentences. In this way, everyone using it would send similar data, enabling ProtoBot to give better responses."

We observed that as ProtoBot becomes better trained and supports a wider range of hardware, it can facilitate the creation of more reliable designs. Moreover, we saw the potential for a ProtoBot that generates code more quickly and accurately to assume the role of a guide that simplifies design processes without interfering with design decisions, rather than merely serving as an interactive conversational tool.

In the future, tools like ProtoBot could offer a new way for people to use AI, both in selecting which electronic components to use during the design ideation phase and in preparing prototypes after obtaining the components. We predict that advancements in AI will not only affect how electronic components can be used and designed but also influence how they are presented to users in marketplaces. We foresee that people will be guided to design under the reliable assistance of artificial intelligence.

\subsection{Adjusting Design Parameters When Interacting With the Prototypes}




Another key lesson we learned is the importance of allowing users to adjust design parameters when interacting with the prototypes. P4 expressed a desire for the ability to change and experiment with the threshold value of the algorithm created to distinguish between the stepping motion and the action of covering the litter in the FitFit application, so that she wouldn't have to request new code for each different threshold.

Similarly, P3 highlighted that, due to their limited knowledge in the area, the parameters set automatically by ProtoBot created confusion in the process. They wished for an interface that would allow them to adjust settings and experiment without having to re-enter prompts. The participant stated, "Because I am not very knowledgeable in this area, the parameters it set on its own confused me. I wish I didn't have to give prompts to change some settings and make experiments. Instead, it could offer me an interface alongside the code it uploads to the device, where I can adjust parameters. For instance, if it's going to show a bar indicating temperature measurement while the heat rises, it could allow me to set how much the bar fills within which temperature range by experimenting."

As a result, we identified a clear consensus that in real-life scenarios, users regularly need to fine-tune certain parameters without making significant changes to the prototypes. This need frequently arises especially after the initial prompts. We realized that this adjustment can be facilitated by providing users with a space to modify parameters or experiment without requiring them to input new prompts. By incorporating such features, ProtoBot can offer a more flexible and user-friendly experience, allowing for iterative refinement of prototypes.


We foresee that in the future, tools like ProtoBot will have interfaces that, in addition to text or voice-based conversational interactions, allow users to change the parameters or settings created in the background in other ways. Realizing this led us to consider questions like, "If in the future we communicate with tools like ProtoBot via voice, how could we present these parameters to users?"

\section{Conclusion}


In this article, we presented ProtoBot, an AI-driven chatbot designed to enable individuals with limited software and hardware expertise to prototype wearable electronic devices without the need to write code. By leveraging large language models and integrating with Arduino-compatible hardware, ProtoBot simplifies the prototyping process through prompt-based interaction. Our case study with four professionals from diverse disciplines demonstrated that ProtoBot not only functions effectively but also has a positive impact on reducing the anxiety associated with inexperience in hardware and software design.

The participants were able to conceptualize and prototype their own wearable devices, transforming initial uncertainty into a positive and motivating experience. From our findings, we derived three key lessons: (1) uncertainty can be reframed as a positive driver for exploration, (2) ProtoBot has the potential to act as a reliable guide for users with limited expertise, and (3) users may need to adjust design parameters when interacting with prototypes, highlighting the importance of iterative refinement in the design process.

Our work lays the foundation for helping individuals with limited hardware and software knowledge to explore wearable prototyping possibilities. We believe that tools like ProtoBot can empower non-experts to engage in wearable technology design. As AI algorithms continue to advance, we anticipate that ProtoBot and similar tools can be tailored to handle more complex interactions and support a wider range of hardware components. For future work, we encourage researchers to participate in the creation of making prototyping more accessible, pioneering rapid prototyping without fear of uncertainties for those who wish to develop both wearable prototypes and other products.

\bibliographystyle{ACM-Reference-Format}
\bibliography{sample-base}

\end{document}